\documentclass[a4paper]{ifacconf}

\usepackage{graphicx}      
\usepackage{natbib}        

\usepackage{todonotes}

\usepackage{amsmath,amsfonts}

\usepackage{enumitem} 

\usepackage{balance}

\begin{document}
\begin{frontmatter}

\title{Feedback and Time are Essential for the Optimal  Control of Computing Systems
} 



\author[First]{Eric C.\ Kerrigan} 

\address[First]{Imperial College London, UK, (e-mail: e.kerrigan@imperial.ac.uk)
}

\begin{abstract}                
The performance, reliability, cost, size and energy usage of computing systems can be improved  by one or more orders of magnitude by the systematic use of modern control and optimization methods. Computing systems rely on the use of feedback algorithms to schedule tasks, data and resources,  but the models that are used to design these algorithms are validated using open-loop metrics. By using closed-loop metrics instead, such as the gap metric developed in the control community, it should be possible to develop improved scheduling algorithms and computing systems that have not been over-engineered. Furthermore, scheduling problems are most naturally formulated as constraint satisfaction or mathematical optimization problems, but these are seldom implemented using state of the art numerical methods, nor do they explicitly take into account the fact that the scheduling problem itself takes time to solve. This paper makes the case that recent results in real-time model predictive control, where optimization problems are solved in order to control a process that evolves in time, are likely to form the basis of scheduling algorithms of the future. We therefore outline some of the  research problems and opportunities that could arise by explicitly considering feedback and time when designing optimal scheduling algorithms for computing systems.
\end{abstract}


\end{frontmatter}

\section{Optimal Computers Use Control}

\label{sec:stateoftheart}



Computing systems may be composed of reliable and efficient \emph{components},  but
 reliability of the overall \emph{system} comes at the cost of  large inefficiencies due to over-engineering.  In embedded computing applications, such as avionics,  engineers are  constantly seeking to reduce the size, weight and power 
of the   systems, which results in significant savings in energy, cost, maintenance and improved safety. 
 In traditional data centres, for example, a server can  draw 70--90\% of its maximum power   when it is not doing any work.  Information and Communication Technologies (ICT) were  responsible for producing at least 7\% of worldwide electricity consumption in 2008 and this figure 
is expected to rise to more than 14\% by 2020~\citep{vereeckenetal:2010}.
Though the GeSI SMARTer~2020 report~\citep{gesismarter2020}
 claims that ICT-enabled solutions has the potential to reduce greenhouse gas emissions by 16.5\% in 2020, there is clearly a  need for computing systems to reduce their own  contribution to these emissions. 
 
 What is lacking, however, is a complete theory that allows  engineers to understand  how  various components interact with each other and what effect this  has on the overall system behaviour.  By building on recent results from control theory and mathematical optimization, disciplines where \emph{feedback} enables engineers to design  robust and efficient systems, it is possible to design computing systems that can be one or more orders of magnitude more energy efficient, cheaper, faster, smaller and  reliable than today. 

Every computing system today employs \emph{feedback}  in some form or other to guarantee a certain level of performance and reliability in the presence of \emph{uncertainty}, such as unpredictable work-loads, computational and communication delays, data losses    and component failures. 
Problems that require feedback  algorithms arise in a variety of contexts in computing systems~\citep{hellerstein:dao:parekh:tilbury:2004,hellerstein:singhal:wang:2009}:
\begin{itemize}
\item
Data, tasks and resources (such as   processors, storage and communication networks) need to be managed to achieve  a certain quality of service, 
guarantee that computations are correct and ensure that tasks  are completed before  deadlines. 
\item  Minimization of  power consumption  and   overheating protection by dynamic voltage and frequency scaling 
and smart scheduling of jobs.
	\item Estimating  the workload and available resources, as well as the status and completion rate of jobs.
	\item Guaranteeing resilience of the system in the presence of faults and cyber attacks.
\end{itemize}

\begin{figure}[tb]
\centering
\includegraphics[width=\columnwidth]{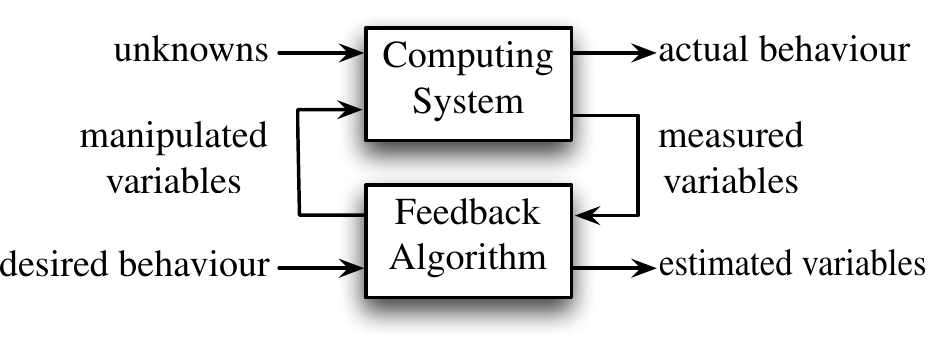}
 \caption{Feedback-based computing system.}
 \label{fig:feedback}
 \end{figure}

Figure~\ref{fig:feedback} shows the key components of a typical feedback-based computing  system. 
 Unknowns, such as  future workloads and  data losses, act on the system. It is often possible to measure  key variables, such as  power consumption, duration  of a computation or resource utilisation. The feedback algorithm might use these measurements  to update  a  model of the computing system to correct for errors in the estimates of the measurements. If the actual behaviour of the system is different from the desired behaviour, the feedback algorithm   updates the values of certain manipulated variables, such as the processor clock frequency or memory allocation, until the behaviour is as desired or the estimates are sufficiently accurate.

\subsection{Why is the control of computing systems challenging?}

Computing systems present a number of significant challenges that  stretch  the theory and practice of  modeling, control and optimization well beyond what is  possible with the state of the art:
\begin{itemize}
\item Time-scales at which the dynamics evolve range from pico-seconds to hours and even days. In some applications the speed of a feedback algorithm is not important and in others absolutely critical.
\item Power consumption can range from tens of MW to less than a pW. The power consumed by the feedback algorithm itself may  therefore have to be  minimized.
\item The cost of a system can range from a few cents to billions of euros. The silicon area and cost of implementing a feedback algorithm may therefore be absolutely critical to the application.
\item The same system may have to execute a variety of tasks with mixed  criticalities. Some tasks or units are not allowed to miss their deadline or fail, whereas delays and data losses are acceptable in others.
\item Demands by an application on the speed or resources can vary by orders of magnitude in a single run. Uncertainty models based only on the worst-case or  average-case may therefore not be practical.
\item The number of computing units that may interact with each other and the number of tasks may vary from one to billions or more. Feedback algorithms therefore need to be scalable.
\item Processing units can all be located on one chip or spread across the world. Feedback algorithms therefore might have to be implemented in a decentralized manner, while  guaranteeing correct overall system behavior.
\item Computing systems have hybrid dynamics. Methodologies are  needed that can cope with the interaction between discrete dynamics on the computation side, e.g.\ logic and discrete states or events, and continuous dynamics on the physical side, e.g.\ heat dissipation or energy usage.
\item First-principles modeling of computing systems is still very much in its infancy. A mixture of new first-principles and data-driven modeling methodologies needs to be developed.
\end{itemize}

 \subsection{Not all computers are on desktops or in data centres}
 \label{sec:mobile_sensors}
 
 It is often the case that embedded computing systems are mobile and/or part of a sensor network. 
Typical examples include: (i)~automotive electronics or avionics systems, (ii)~mapping of traffic or  pollution in a city, (iii)~automated manufacturing, farming or warehousing, or (iv)~UAVs flying in formation to reduce air drag,  fighting forest fires or performing remote earth sensing tasks. 

In  these applications there is often a need for  individual  nodes to cooperate towards satisfying high level tasks, which require a significant amount of computation power. Consider the fire fighting scenario, for example. The  system has to use a numerical model of the fire dynamics to predict how the fire will develop, solve an optimization problem to determine where and when to send each UAV and fire fighting unit, as well as perform local control of each UAV. Because of the unpredictable environmental conditions, the system should be fault tolerant and  be able to do  the simulation,  data processing, coordination and planning  by itself in real-time, by combining the processing power of each  node in the network, rather than sending all the data to a central high performance computing facility. 
 
Mobile computing systems and sensor networks present many of the challenges to modeling and  control  mentioned above, with the addition that the  quality and  structure  of the communication network varies with time, hence the topology of the computation network  has to change. If the nodes are mobile, then  the position and propulsion energy  also have to be integrated when determining how best to control the computation, communication and data storage.

The main point to note here is that the computing system often interacts with the physical world and/or vice versa in real-time. Traditional methods for the control of computing systems do not always explicitly acknowledge  or take advantage of this fact. It therefore makes sense to consider the modeling and control of the combined cyber-physical system, 
rather than treating the computing system as separate from the physical system.

\subsection{Where are we?}

One of the   reasons why computing systems are over-engineered is because  feedback algorithms in computing systems are usually designed in an ad hoc manner without  systematic use of methods  from the rich body of control theory,  the science of feedback in dynamical systems. 
This is often also not helped by there being some slight, but important, differences in terminology  between the computing and control  communities. In computing,  the terms `dynamic' or `static' are used where a control engineer would have insisted on using  `feedback/closed-loop' or `open-loop', respectively. In control theory, feedback algorithms can be dynamic or static.  Sometimes `dynamic' or `static' are used in the computing literature when, respectively,  `open-loop time-varying' or `open-loop constant' would have been consistent with the control theory literature.     A control engineer would usually agree that a computer engineer's `feedback scheduler'  is a  feedback algorithm.

A number of academic  and industrial research groups have reported significant improvements in the response times, quality of service, reliability, energy and resource usage of computing systems  by systematically implementing control-theoretic feedback algorithms in the design of their computing and software systems~\citep{hellerstein:dao:parekh:tilbury:2004,hellerstein:singhal:wang:2009,sha:historical:2004}. 

Research in this area is still very much at an early stage.
Many  existing techniques for the control of computing systems 
are mostly based on control theory that was state of the art in the 1980s. Controller design techniques that have been implemented range from classical PID control to robust control  using $\mathcal{H}_\infty$ and LQG design. 

Over the last two decades, however, there have been major developments in modeling  for control~\citep{LeeSeshia2014,GoebelSanfeliceTeel2012,hjalmarsson:2005} and optimization-based  control, often also called model predictive control~\citep{christofidesetal:2013,mattingley:wang:boyd:2011,negenborn:2014,mayne:2014}.  


\subsection{Where can we go?}

It should be possible  to introduce a step change in the design of computing systems by building on  recent advances in control and optimization in order to  develop:
\begin{enumerate}[label=\alph*)]

\item \emph{Modeling techniques that capture  dynamics critical to  feedback algorithm design.} Within the computing  community, \emph{open-loop} models are   used to assess the quality of a model before  designing feedback algorithms, followed with extensive closed-loop simulations and many  design iterations before implementation~\citep{hellerstein:singhal:wang:2009,hellerstein:dao:parekh:tilbury:2004}.  
It might be productive to take a   different, more sophisticated  approach developed in the control community, and use ideas similar to the \emph{gap metric}~\citep{georgiou:smith:1997,vinnicombe:2001,lanzon:papageorgiou:2009} for determining whether two   systems are similar in \emph{closed-loop}.

\item  \emph{Real-time optimization-based scheduling algorithms.}
Scheduling problems are most naturally posed as constraint satisfaction or mathematical optimization problems~\citep{hellerstein:singhal:wang:2009}, but  they have traditionally not been solved using numerical optimization methods~\citep{li:wu:2013,sha:historical:2004,davis:burns:2011}.  
Optimization methods could instead be used to solve scheduling problems in real-time by building on recent, computationally efficient real-time model predictive control methods~\citep{bemporadetal:2015,domahidi:zgraggen:zeilinger:morari:jones:2012,diehl:bock:schloder:findeisen:allgower:2002,zavala:anitescu:2010,zavala:biegler:2009}.
It should also be possible to use recent results from cooperative model predictive control~\citep{christofidesetal:2013,negenborn:2014} to develop  scheduling algorithms that enable  distributed, multi-processor computing systems  to cooperate in meeting overall system specifications. 
\end{enumerate}

This paper discusses the above two topics  in detail. Whereas the main focus of this paper is on computing systems, many of the points raised below are equally valid for control design in other application areas, such as transport, buildings, manufacturing, energy and healthcare.

 \section{Mind the Gap}

\label{sec:gapmetric}

The implementation of control-theoretic methods to computing system design has largely been hindered by the difficulty in obtaining sufficiently accurate  dynamical models~\citep{hellerstein:singhal:wang:2009,hellerstein:dao:parekh:tilbury:2004}. This gap can be bridged by bringing state of the art modelling techniques from the control theory literature to the  computing community. Likewise,  computing systems  present unique  challenges that will require new modeling techniques and associated  numerical methods to be developed by the control and optimization  communities.

\subsection{When are two different systems similar?}


When designing a feedback algorithm, it is  important to remember that {a good open-loop  model is not  necessarily  a good model for  feedback algorithm design. Likewise, a good model for feedback design is not necessarily a good open-loop model}.

\begin{figure}[!tb]
\centering
\includegraphics[width=\columnwidth]{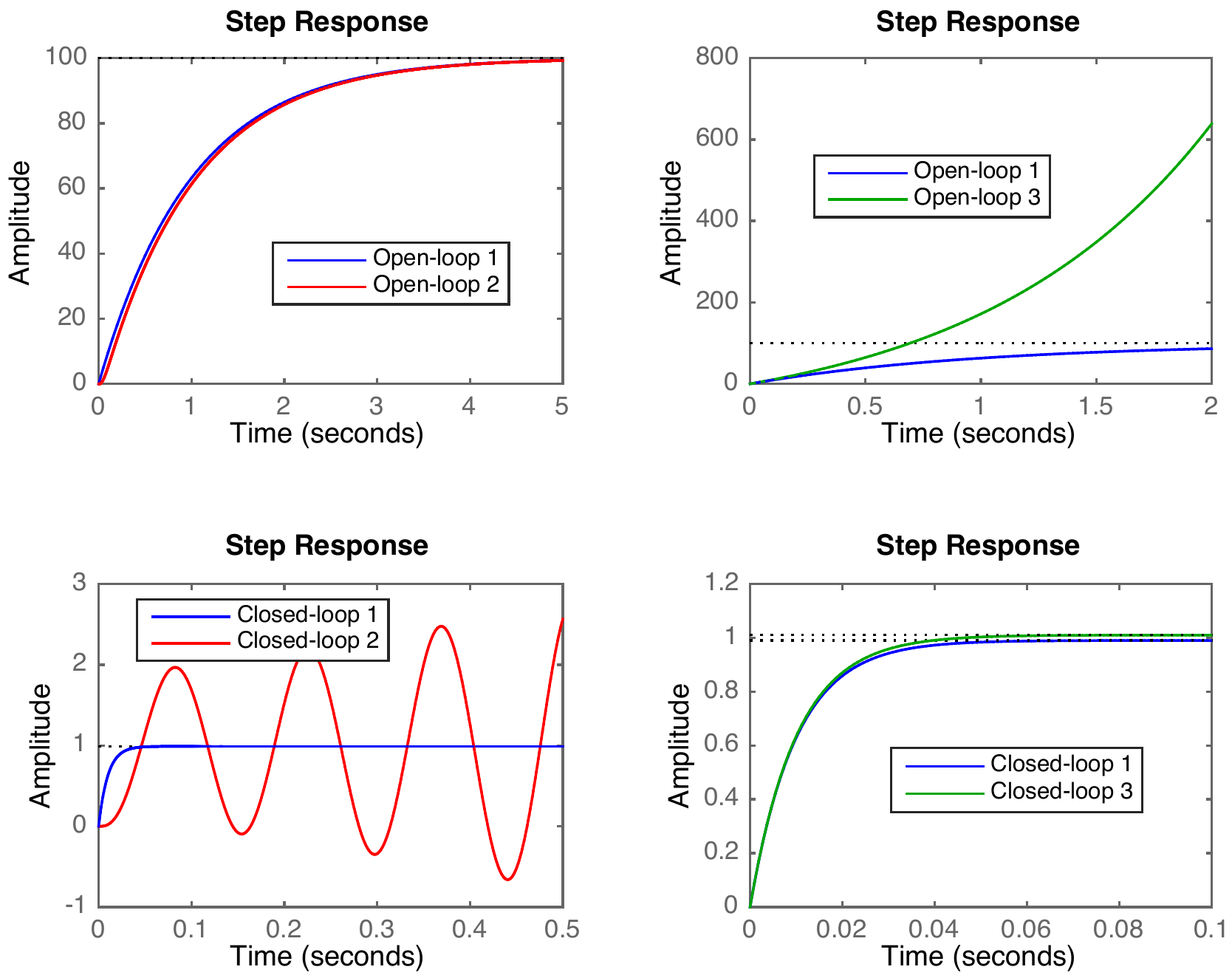}
 \caption{\citep[Fig.~12.3]{astrom:murray:2008} Similar open-loop responses do not imply similar closed-loop responses, and vice versa. 
 See \citet[Sect.~12.1]{astrom:murray:2008} for  details of the  models,  feedback policy and gap metric computations.
 }
 \label{fig:closedandopen}
\end{figure}

The two plots on the left  of Figure~\ref{fig:closedandopen}
 show that a good open-loop model is not sufficient for feedback design. The systems  have similar open-loop responses, but  different closed-loop responses. This is because  the controller  amplifies the effect of differences in open-loop dynamics, resulting in an unstable closed-loop for one system. Equally surprising, the two plots on the right  of Figure~\ref{fig:closedandopen} show that   systems can have  very different open-loop responses (one system is open-loop stable and the other is open-loop unstable), but   virtually indistinguishable closed-loop responses. 
Feedback therefore allows one to tolerate  large uncertainties in certain cases (the right in Figure~\ref{fig:closedandopen}), without any significant difference in closed-loop performance. However, in other cases it is critical to  capture the dynamics where a poorly designed feedback algorithm could  make matters worse  (the left in Figure~\ref{fig:closedandopen}).  A good understanding of control theory, coupled with appropriate model development, is  therefore necessary  to avoid designing computing systems that are over-engineered, as is the case today.

The above example partly  explains why it has been  difficult to use existing models from the computing systems literature when designing feedback algorithms. Queuing theory has been  very successful in modeling the steady-state behavior of computing systems and networks, but has not  yet been so successful in modeling  the transient behavior~\citep{hellerstein:dao:parekh:tilbury:2004}. Nonlinear fluid models have  been widely used in modeling for  network congestion control, but have not been so successful in the modeling of computing systems, because workloads have  complicated characteristics, architectures are multi-tiered  and  the nature of the limiting resource changes with time~\citep{hellerstein:singhal:wang:2009}. Uncertainties in computing systems and networks are usually modeled as additive process noise, but  this  structure does not allow a stable and an unstable model to be compared. Likewise,  parametric uncertainties cannot be used to model  differences in the model order or   time delays, especially if the delays change with time or are state-dependent.

It should be possible to extend a very sophisticated concept, developed in the control community,  to the modeling of computing systems. Instead of using open-loop metrics, one could use  the  \emph{gap metric} for determining  whether two   systems are similar in \emph{closed-loop}~\citep{vinnicombe:2001,georgiou:smith:1997,lanzon:papageorgiou:2009,james:smith:vinnicombe:2005}. Loosely speaking, the gap  from one  system to another is  the   size of the  smallest  dynamical system  that  needs to be connected (defined in an appropriate sense) to  the first system  in order for the input-output responses of both  systems to be the same. 
If two systems have a small gap between them and the same feedback algorithm is used, then  it is possible to guarantee  that  the  robustness and performance of the two closed-loop systems will be  similar,  as  in the right half of Figure~\ref{fig:closedandopen}. If the gap between two systems are large,  but the open-loop responses are similar, then  it is possible that the closed-loop systems will behave very differently, as in the left half of Figure~\ref{fig:closedandopen}. 

The gap metric allows one to  account for dynamic and parametric uncertainty, additive disturbances as well as unstructured uncertainty, which greatly reduces the difficulty of modelling the uncertainty and hence the design of a  feedback algorithm. 
 It is also possible to account for uncertainty in the number of unstable  modes or system zeros, which impose fundamental performance limitations on the design of  feedback algorithms. 

It will therefore be of interest to investigate whether the gap metric can be used to develop  and validate first-principles and data-driven models for  computing systems. 
This research could therefore bring about a fundamental change in the way that computing systems are modeled and feedback algorithms are designed.

\subsection{Use your brain}

As a first step,  it will be useful to revisit first-principles  methods  currently used in modeling computing systems, such as queuing theory and  linearized fluid flow models, but armed with the gap metric and associated  control design tools. These methods could be used to analyze the robustness of existing modeling approaches  and the fundamental limitations for controller design. However, it might be necessary to extend both the state of the art in modeling  computing systems as well as   develop techniques that are new to the control community.

Recent research on the control of infinite-dimensional systems~\citep{jones:kerrigan:2010, jonesetal:2015} has shown that, compared to using open-loop metrics, careful use of {the gap metric allows one to: (i) shorten the design phase, (ii) synthesize feedback algorithms  that are computationally less demanding and (iii) have stronger guarantees on the robustness and performance of the  closed-loop system}. 
One way in which the ideas in~\citet{jones:kerrigan:2010} could potentially be applied in the control of computing systems is to approximate time delays, which are infinite-dimensional systems, with finite-dimensional input-output models and use the gap metric to provide guarantees on the closed-loop behavior. The use of Pad\`e approximations to model time delays is a standard technique in control engineering, but does not yet seem to have found widespread use in the computing community. The gap metric can also allow one to provide   robustness guarantees on the closed-loop behavior  if the delay is uncertain~\citep{cantonietal:2012}.

An important   aspect to consider in the modeling of computing systems, where research is  in its early stages, is how the \emph{physics} of computing systems should be incorporated into models. It is often important to consider the power consumption, heat dissipation and  dynamics of the cooling system.  As discussed in Section~\ref{sec:mobile_sensors}, in many cases the nodes in a distributed computing system are mobile and the physical environment could affect the  performance of the network. Due to the interaction between sub-systems with discrete states, events, logic and sub-systems with continuous states and dynamics, computing systems are therefore best modeled as cyber-physical systems using techniques from \emph{hybrid dynamical systems theory}~\citep{LeeSeshia2014,GoebelSanfeliceTeel2012}.

A major open question that has to be addressed is whether nonlinear gap metric ideas~\citep{georgiou:smith:1997,james:smith:vinnicombe:2005} can be extended to certain  types of  hybrid systems, while  allowing one to compute bounds on  the gap  between two systems. Furthermore, even if the model is linear, the optimal control policy is nonlinear, in general. As a consequence, there is considerable scope for extending nonlinear gap metric results for the analysis and design of model predictive controllers.



\subsection{Use the data}

Because of the lack of first principles models, researchers in computing systems often apply well-established, open-loop stochastic  system identification methods~\citep{ljung:1999} to develop  models from input-output data, with some success~\citep{hellerstein:dao:parekh:tilbury:2004}.   As a starting point for future research, it would be of interest to compare the methods currently used in the computing systems literature against recently developed  methods for model validation, identification and parameter estimation of linear, nonlinear and hybrid systems~\citep{hjalmarsson:2005,ljung:2010,paoletti:juloski:ferraritrecate:vidal:2007,betts:2010}.  

However, as discussed above, it is critical to bear in mind that closed-loop measures should  be used when validating or identifying models for feedback algorithm design. 
Gap metric ideas can also be used  if  input-output data  is available to validate  a given model~\citep{hjalmarsson:2005}  or to  compute a model directly from  data using system identification methods~\citep{date:vinnicombe:2004}. Given some data, a model can be interpreted as sufficiently accurate if the data is consistent with what one would have measured if a small dynamical system is connected (defined in an appropriate sense) to the model. There are no published results on how closed-loop metrics can be used for model validation and identification of computing systems. Research in this area could therefore   enable the development of more appropriate models and feedback algorithms, with better performance and robustness guarantees than methods based on open-loop measures. 

Due to the complicated nature of the dynamics,  large size and high speed of computing systems, state of the art  optimization methods will be inadequate  for solving many of the new  model validation, identification and parameter estimation problems that will arise. Hence, novel and more efficient numerical methods will have to be developed that allow one to  (in)validate, (un)falsify or identify a feedback-oriented model, ideally in real-time.

\section{It Takes Time}
\label{sec:suboptimal}

Because time does not stop, an approximate answer today can be better than an accurate answer tomorrow. 
Computing systems employ feedback algorithms  to cope with uncertainty, but the system is in open-loop while the computation is being carried out. 
Hence, it might  be better to implement a simple, computationally efficient algorithm at a fast rate than a sophisticated algorithm at a slow rate.  

\subsection{Computing needs time}

\begin{figure}[tb]
\centering
\includegraphics[width=\columnwidth]{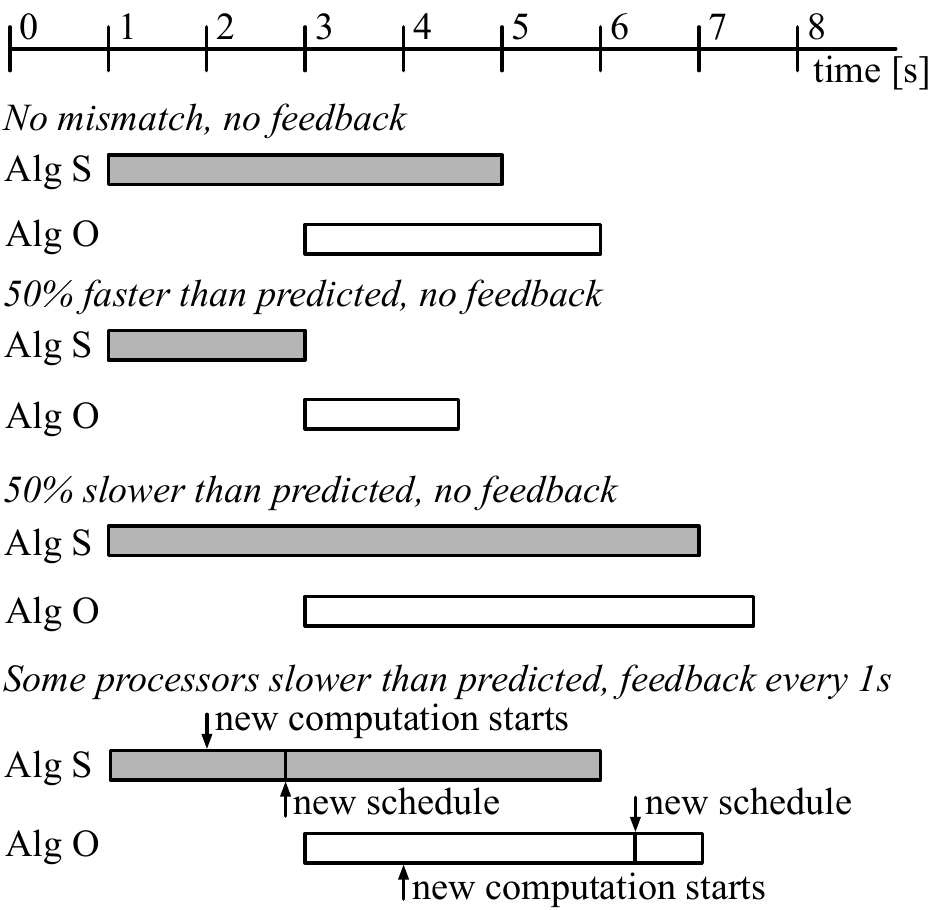}
 \caption{Comparison between two scheduling algorithms. }
 \label{fig:suboptimal}
 \end{figure}

Consider a  simple  example, illustrated in Figure~\ref{fig:suboptimal}. 
Suppose you have a  large number of  jobs that need to be completed, a large number of heterogeneous processors to do the computations and two scheduling algorithms.  Algorithm~O is the `optimal' algorithm, which is guaranteed to find the global minimum, and Algorithm~S is the `sub-optimal' one, which is not guaranteed to find the global minimum or even a local minimum.  Algorithm~O takes 3 seconds to compute a schedule, but  Algorithm~S only takes 1 second. Suppose there is   no mismatch between the  speeds of the processors and those assumed by the algorithms. At the first scheduling event,  at time $t=0$\,s,  both algorithms start to compute a schedule and Algorithm~S terminates after 1\,s with a schedule that will take 4\,s from start to finish, i.e.\ all jobs will be completed by $t=5$\,s. However,  Algorithm~O terminates after 3\,s with a schedule that will take  3\,s from start to finish, i.e.\ all jobs will be completed by $t=6$\,s. In this case,  `sub-optimal' Algorithm~S is better than  `optimal' Algorithm~O, because Algorithm~S gets all the jobs done before Algorithm~O! Algorithm~O needs to take less than 2\,s to compute a solution in order to be better than Algorithm~S.

The difference can be even greater if there are mismatches between the  speeds of some of the processors and those assumed by the algorithms, as   is also shown in  Figure~\ref{fig:suboptimal}. Suppose that the processors are faster than assumed and that it  takes 50\% less time for jobs to complete than those predicted according to  the original schedules, i.e.\  the schedule computed by Algorithm~S completes all jobs by $t=3$\,s, which is when Algorithm~O has just completed its computation (the difference between  completion times is also larger). Consider the opposite case in which    processors are much slower than assumed so that the original schedules take 50\% longer than predicted. However, suppose now that the job completion rate of each processor is fed back every~1\,s. Algorithm~S can detect  this information at $t=2$\,s and implement a new schedule  at, say, $t=2.8$\,s (there are fewer uncompleted jobs, hence  it takes less time to compute a schedule). It is therefore possible that Algorithm~S can  use the updated completion rates to find a new schedule that will get all remaining jobs completed by $t=6$\,s. However, Algorithm~O would  detect the actual job completion rates at $t=4$\,s and might only be able to implement  a new schedule for the remaining jobs from, say, $t=6.3$\,s, which is after the schedule computed by Algorithm~S would have completed.

This example demonsrtated that  a scheduling algorithm should ideally take into account the time  it takes  to compute a schedule \emph{and} that  feedback helps reduce the effect of incorrect assumptions. Most  algorithms and abstractions in computing do not explicitly take the passage of time into account. Real-time operating systems are arguably not as real-time as they should  be. Computing not only takes time, but   needs time~\citep{lee:2009}. 

\subsection{When is an optimal scheduler not optimal?}

Data, tasks, processors, networking and storage  need to be scheduled to meet  deadlines and achieve  a certain quality of service, while minimising energy usage. Scheduling problems are most naturally posed as constraint satisfaction or mathematical optimization problems. 
Furthermore, many well-known scheduling algorithms 
are  feedback algorithms. 
Scheduling problems are   perfect candidates for combining solutions from control and optimization theory.  

Because the  resulting  optimization problems can be computationally intractable, scheduling problems have historically not been solved using numerical optimization methods. 
Instead, computing researchers have developed a  range of  computationally efficient or heuristic strategies, which are usually expressed as  simple sets of rules.

In the hard real-time scheduling literature, an algorithm is often defined to be optimal if the algorithm can schedule all task-sets that can be scheduled by any other algorithm. Under very specific and often conservative assumptions on the task-sets and computer architecture, it can be shown that certain well-known classical scheduling algorithms, such as earliest deadline first 
or rate monotonic, 
are optimal in this sense.  There also exists a vast array of sub-optimal scheduling algorithms and in some cases  one can  compute  limits on the level of sub-optimality.

In many applications it makes sense to relax some hard  timing constraints and replace them with soft  timing constraints, where the aim is to minimize the violations. 
For example, in video conferencing the difference
between worst-case and average bandwidth requirements  can be more than one order of magnitude~\cite[Sect.~4.1]{sha:historical:2004}. A user might be willing to tolerate an occasional delay or data loss, or the processor frequency could be reduced to save on power   requirements. Hence, if one were to use existing hard real-time scheduling algorithms, then the computing system might be over-engineered by most  measures, such as cost, energy or speed.  
There is therefore significant scope   for improving system performance measured according to  criteria other than hard time deadlines.  Research could therefore be devoted to   developing new methodologies to solve practical scheduling problems for which the very restrictive conditions on the task-sets and architectures, currently assumed in the classical real-time scheduling literature, can be relaxed. 


 There have been dramatic improvements in numerical optimization methods over the last few decades. 
The last few years have therefore seen a sharp increase in the use of numerical optimization methods for solving scheduling problems, mainly driven by the need to minimize energy usage in data centers and energy-limited computing devices~\citep{li:wu:2013}. However, much of the literature either (i)~focuses on steady-state optimization, hence ignoring transients, (ii)~uses open-loop dynamic models, (iii)~do not explicitly account for the time taken to solve the scheduling problem, (iv)~do not  consider the effect of terminating the optimization solver before a solution has been found, or (v)~only consider hard time constraints.

 \subsection{Any time in real-time  predictive control}
 
 There is a clear need to develop new scheduling algorithms and abstractions that explicitly address the passage of time. This can be done by incorporating into the algorithm a time-based  dynamical model of the system and uncertainties, regularly updating the algorithm with the current state of the resources and completion rates,   and  implementing the best available solution  when  further computations are not guaranteed to improve the  closed-loop system performance or robustness.

It is therefore possible to take a different approach to most scheduling methods and build on recent research in efficient real-time algorithms~\citep{bemporadetal:2015,domahidi:zgraggen:zeilinger:morari:jones:2012,diehl:bock:schloder:findeisen:allgower:2002,zavala:anitescu:2010,zavala:biegler:2009} and computer architectures~\citep{jerez:ling:constantinides:kerrigan:2012,jerezetal:2014} for optimization-based control. A dynamical model of the system can be used to formulate an optimization problem, which is updated at each sample instant with the latest measurements and solved using numerical optimization methods before implementing the first part of the solution. This process is then repeated at all sample instances. However, the key idea in  real-time model predictive control algorithms is that one does  \emph{not} iterate till the algorithm has converged, but that the algorithm is allowed to  terminate at \emph{any time} with a potentially sub-optimal solution. 

 At each  scheduling event, the optimization solver can be initialized with a version of the  policy obtained at the previous event that is time-shifted, as in receding horizon control, or truncated, as in decreasing horizon control. 
Using similar arguments as in the real-time predictive control  literature one should be able to construct optimization-based schedulers that will  converge to a locally optimal solution after a few scheduling events, provided feedback occurs at a sufficiently fast rate.  In order to take advantage of any existing and future results in the real-time  scheduling literature, one can also choose to initialize the optimization solver with  the policy that one would get from implementing any other rule-based or heuristic scheduling algorithm. 

An anytime approach requires significantly less computational resources than iterating till an optimal solution has been found.  
Real-time model predictive control can allow one to implement sub-optimal solutions at a fast rate with  similar or better closed-loop performance, coupled with   a significant reduction in computational requirements, compared to implementing optimal solutions at a slow rate.

Provided the right algorithms and computer architectures are used, optimization-based controllers can  be implemented for very fast systems with sample rates in the MHz~\citep{jerezetal:2014} range, and has is sufficiently efficient for controlling the speed and power dissipation of  microprocessors~\citep{mattingley:wang:boyd:2011,zaninietal:2013}. It is clearly time for real-time model predictive control of  computing systems.

\subsection{Computers working  together}

In many computing systems today each processing unit functions in a non-cooperative, decentralised manner. Though this  has allowed for the massive expansion of the Internet, this approach is not always ideal or necessary.  Many high performance and embedded computing systems have custom-designed architectures and operating systems that allow the processing units to share information and resources in an effective and reliable manner. On the other hand, it is also not always sensible  to have a purely centralised approach either, since expansion is difficult and the system can be  more vulnerable to  faults and attacks. A  compromise between a fully centralised and fully decentralised  approach is a {cooperative distributed design}, where 
computing units  share  information and resources with a common goal. 
Most well-known scheduling algorithms are applicable to uni-processor systems only 
and research on   multi-processor and distributed architectures is still very much in its infancy~\citep{davis:burns:2011,li:wu:2013,vidyarthi:sarker:tripathi:yang:2009}. 
There is therefore a need for  research on scalable, hard and soft real-time scheduling algorithms for distributed computing systems.

Over the last decade there has been an explosion of activity  in the control community in the area of distributed control
of  networks of dynamical systems. This activity has  resulted in the development of scalable, real-time optimization-based control methods tailored to cooperative  distributed systems~\citep{christofidesetal:2013,negenborn:2014}. These  
methods could, in principle, be applied to develop scalable scheduling algorithms that enable  distributed, multi-processor computing systems  to cooperate in meeting overall system performance and reliability specifications. 


An open problem is  how best one could develop tractable methods for obtaining low-order models of physically distributed computing systems. The problem with most model reduction methods~\citep{antoulas:2005} is that they require a high-order model. The gap-metric based approach in  \citet{jones:kerrigan:2010} is fundamentally different and does not require  a high-order model, hence is computationally more efficient, while still providing guarantees on the robustness and performance of the closed-loop system. By gradually increasing the model complexity, convergence of the model sequence  happens  faster with closed-loop metrics than with open-loop metrics.  Another advantage is that the resulting model retains the structure and sparsity of the original. This structure can  be exploited by numerical algorithms for design and implementation, whereas most model reduction methods  destroy  structure and sparsity. 
It might therefore be possible to use gap metric  ideas to produce scalable,  distributed models and account for  the effect of communication faults and delays, changes in the structure of the  communication and computation networks, as well as variability in  resources. 

\subsection{When is tailor-made not a luxury?}

As illustrated in Figure~\ref{fig:suboptimal},  the computational resources used by an algorithm has to be sufficiently small. 
Most off-the-shelf optimization solvers are not able to exploit the special structure that is present in  scheduling problems. Therefore, tailor-made methods  have to be developed that are better than the state of the art by exploiting any structure that is present in the scheduling problem. 

The structure can be exploited by
formulating the scheduling problem as a multistage optimization problem, as is done in optimization-based  control~\citep{betts:2010}, and solving this efficiently  with the aid of  sparse linear algebra. Efficient mixed-integer optimization algorithms for optimal control of systems with integer decision variables~\citep{sager:2006} could  be explored in this context.
It might also be possible to derive conditions under which the scheduling problem can be formulated as a convex and tractable optimization problem, e.g.\ in~\citet{deschutter:vandenboom:2001}  conditions are derived under which  the optimal control of max-plus-linear discrete event systems can be formulated as a computationally tractable linear program.

In many scheduling problems, it is  natural to  introduce integer variables and solve a mixed-integer program. 
In some cases it might be better to  model the problem as a  continuous  optimization problem without  integer variables. For example, suppose there is a large number of jobs that can be grouped into a relatively small number of subsets. and that there is  a relatively small number of identical processors. The  decision variables in the optimization solver can include the fraction of jobs from each subset that are allocated to a percentage of a processor's time, as in deadline partitioning techniques~\citep{levin2010}. This results in smaller and  `nicer' optimization problems than using binary variables to assign  jobs to   processors. 

Another question is  how best to incorporate deterministic and stochastic uncertainties  into the  formulation of the optimization problem. 
Robust optimization methods~\citep{bental:elghaoui:nemirovski:2009} might  then be used to efficiently solve scheduling problems subject to  uncertainties. 
One of the main ideas that can be explored is how to formulate the scheduling problem as an optimal control problem  where the optimization is over feedback policies~\citep{goulart:kerrigan:maciejowski:2006}, rather than open-loop input sequences. 

In some cases it might be best to  use  parametric programming~\citep{borrelli:2003} techniques to compute an explicit solution to the optimization problem. The scheduling algorithm can then be implemented as a lookup table,  similar to the way classical scheduling algorithms are implemented as a set of rules, but with guarantees of optimality and with more flexibility in the nature of the assumptions on the task-sets and architectures. The disadvantage is that often the size of the look-up table  blows up for large problem sizes. Parametric programming  might therefore  be best suited 
to small-scale computing systems, such as multi-core processors~\citep{zaninietal:2013}. 


Many optimization methods cannot be terminated at any time with a guarantee that  sub-optimal iterates will reduce the cost,  satisfy all constraints or  guarantee closed-loop stability or robustness. Possible solutions that one could investigate including adding constraints or modifying the cost function to enforce cost reduction, constraint satisfaction and closed-loop stability~\citep{bemporadetal:2015,domahidi:zgraggen:zeilinger:morari:jones:2012,scokaert:mayne:rawlings:1999}. The modified algorithm might take longer to converge, but  can be terminated at any time with an improved strategy, with better closed-loop performance and robustness. 


Finally, one should consider what effect  the architecture of the computing system has on the computational requirements.  Field Programmable Gate Arrays (FPGAs) and  Graphical Processing Units (GPUs) can also be used to explore the use of parallelism and custom number representations to reduce the computational requirements. Can the architecture   be designed to allow for the development of better scheduling algorithms? 



%

\section{Opportunities and More Problems}

There is tremendous opportunity for control and optimization to make a big impact in the area of computing systems. By combining gap metric ideas with real-time model predictive control methods to design new scheduling algorithms, one might be able to design computing systems that are at least one order of magnitude faster, cheaper, more energy efficient and more reliable, compared to  using state of the art open-loop models and classical real-time scheduling algorithms. 

Because of the  range in complexity, size and speed of computing systems, there is also a vast array of problems that will challenge control and optimization theory. By  solving some of these problems, new methods will result that can also be applied outside the computing domain, e.g.\ in power, manufacturing, transport and healthcare, where similar control and optimization problems arise.

%
\balance
{\small
\bibliography{NMPC2015}             
}                                                     


\end{document}